\documentclass[final,5p,times,twocolumn,draft]{elsarticle}

\usepackage{amssymb}
\usepackage{amsmath}
\usepackage{bm}
\usepackage{mathrsfs}
\usepackage{textcomp}
 \usepackage{placeins}
\usepackage{comment}
\usepackage{natbib}
\biboptions{sort&compress}
\usepackage{setspace}

\usepackage[dvipsnames]{xcolor}

\renewcommand{\eqref}[1]{\textup{{\normalfont Eq.~(\ref{#1}}\normalfont)}}

\usepackage{ulem}
\usepackage{xcolor}

\journal{Extreme Mechanics Letter}

\begin{document}

\begin{frontmatter}

\cortext[cor1]{Corresponding author. E-mail address: sano@mech.keio.ac.jp (T. G. Sano).}


\title{Curling morphology of knitted fabrics: Structure and Mechanics}

\author[a]{Kotone Tajiri}
\author[b]{Riki Murakami}
\author[b]{Shunsuke Kobayashi}
\author[b]{Ryuichi Tarumi}
\author[a,c]{Tomohiko G. Sano{$^{*}$}}

\address[a]{School of Integrated Design Engineering, Graduate School of Science and Technology, Keio University,3-14-1 Hiyoshi, Yokohama, 2238522,Japan}
\address[c]{Department of Mechanical Engineering, Faculty of Science and Technology, Keio University,3-14-1 Hiyoshi, Yokohama, 2238522, Japan}
\address[b]{Graduate School of Engineering Science, Osaka University,1-3 Machikaneyama, Toyonaka, Osaka, 5608531, Japan}


\begin{abstract}

Knitted fabrics are two-dimensional-like structures formed by stitching one-dimensional yarn into three-dimensional curves.
Plain stitch or stockinette stitch, one of the most fundamental knitting stitches, consists of periodic lattices of bent yarns, where three-dimensional (3D) curling behavior naturally emerges at the edges. The elasticity and geometry of knitted fabrics have been studied in previous studies, primarily based on 2D modeling. Still, the relation between 3D geometry and the mechanics of knitted fabrics has not been clarified so far. The curling behavior of knits is intricately related to the forces and moments acting on the yarns, geometry of the unit knitted loops, mechanical properties, and contacts, hence requiring a 3D analysis. Here, we show that the curling of plain knits emerges through the elasticity and geometry of the knitted loops, combining desktop-scale experiments and reduced elasticity-based simulations. We find that by changing the horizontal and vertical knitting numbers (wales and courses, respectively), three types of curl shapes emerge: side curl and top/bottom curl shapes, which are curled only horizontally and vertically, and double curl shape, in which both curl shapes appear together. We uncover that the knit is side-curled when the knitted loop is vertically elongated, while we observe double curl and then top/bottom curl as the loop becomes horizontally elongated. Furthermore, we find that this characteristic loop shape affects the mechanical properties of knitted fabrics. 
The fundamental mechanism of intricate shape deformation is clarified through the force and moment balance along yarn whose centerline shape is discretized through the B-spline curves where elastic stretching, bending, and contact mechanics are taken into account.
We reveal that the 3D structure of the single knitted loop plays a critical role in the curling behavior.
Our results imply that the change in shape per a single knitted loop has the potential to control the 3D natural overall shape of knitted fabrics. The 3D curling behavior of knitted fabrics is useful for industrial applications such as composite materials, wearable devices, and actuators. Our findings could be applied in predicting or designing more complex 3D shapes made of knitted fabrics. 
\end{abstract}


\end{frontmatter}


\section{Introduction}
\label{Introduction}
\begin{figure}[h!]
        \centering  \includegraphics[width=0.48\textwidth]{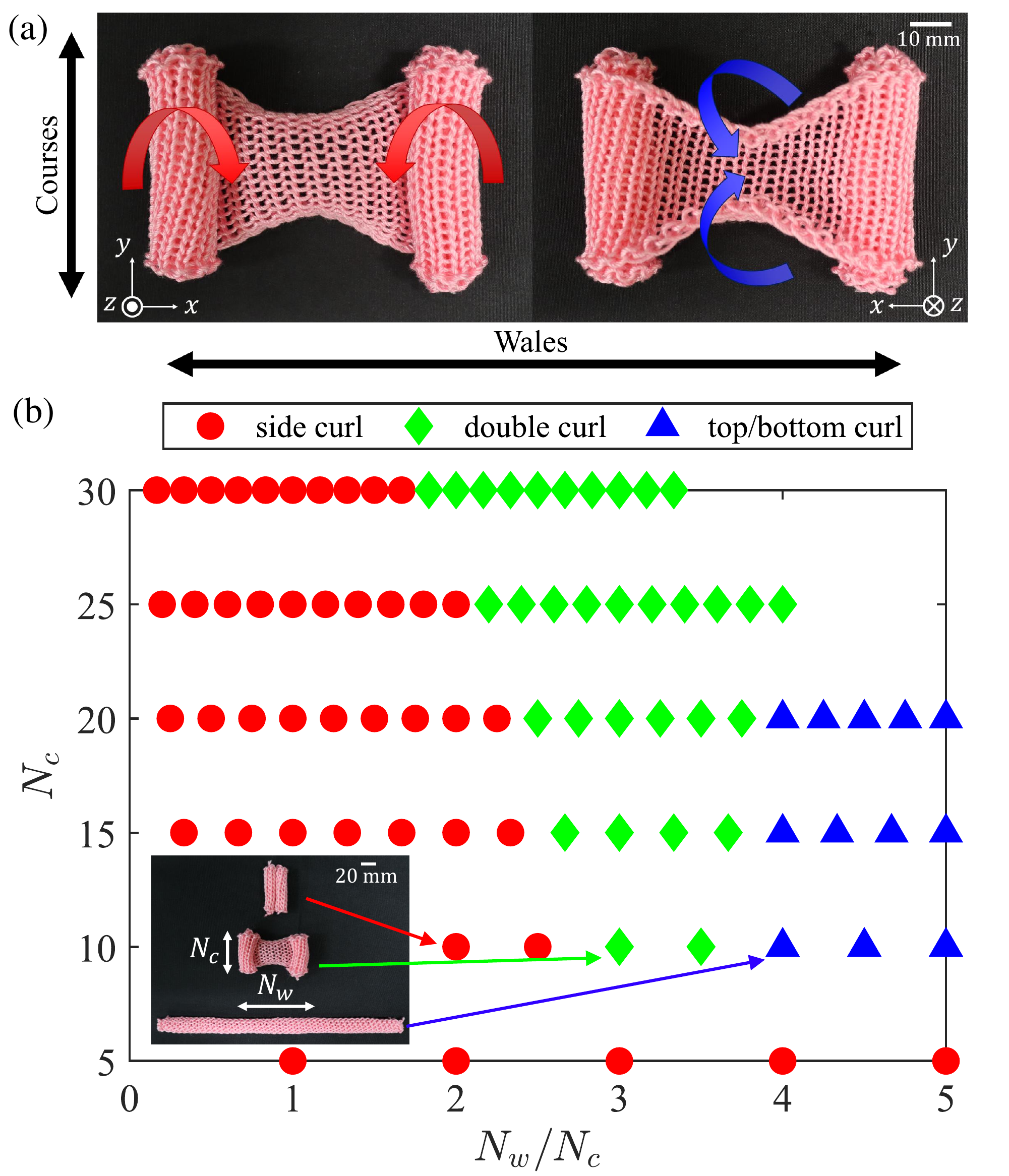}
        \caption{
        Curling morphology of the plain stitches.
        (a) Snapshot of the knit in the double curl state ($(N_w, N_c) = (20, 50)$). The plain stitch curls in two ways; (1)The side curl (left, orange arrows) which curls from front to back in the wale ($x$) direction, and (2)the top/bottom curl (right, blue arrows) which curls from the back to the front in the course ($y$)direction.
        (b)
        Phase diagram of the curling behavior of plain stitches. 
        Three different curling behaviors appear depending on the number of horizontal and vertical stitches ($N_w$ and $N_c$, respectively).
        }
        \label{fig1}
\end{figure}

\indent
A knit is a fabric with a structure consisting of a series of interlaced loops of bent yarn. A plain stitch or stockinette stitch is one of the basic knitting stitches, in which the same loop structure is linked in the horizontal (wale) and vertical (course) directions so that the structure of the knitted loop is the same on the front and back knits with the different stitches between the head and legs of the loop. 
Plain knitted fabrics naturally emerge curls at the edges in two directions. One is the side curl, which curls from the front to the back in the wale direction, and the other is the top/bottom curl, which curls from the back to the front in the course direction~\citep{Bueno2019, Amanatides2022} (Fig.~\ref{fig1}(a)).

Knitted fabrics with an aligned structure of identical loop structures have the property of having large elasticity due to the sliding of the knitted loops under tension. The elasticity of the structure of knitted fabrics is applied to design composite materials, such as increasing material strength by reinforcing the sheet with the elasticity of knitted fibers~\citep{Lim1999, Takano2001}.
The ability to create highly stretchable fabrics is also exhibited in the exceptional draping properties of knitted fabrics and is applied to complex wearable devices~\citep{Haque2016, Luo2022, Paradiso2005}. 
In addition, knitted fabrics
have an anisotropic stretch response attributed to the anisotropic geometry of single knitted loops (bent yarns). Some of the stretchable or bending actuators utilize the anisotropic response of knits~\citep{Albaugh2019, Sanchez2023}.
While the fabrication of 3D shapes has been widely studied, little research has been done on predicting and controlling the curling behavior associated with 3D deformation. Therefore, understanding the 3D characteristics of knitted fabrics and clarifying the mechanisms of the curling behavior is essential for predicting the 3D shape of knitted fabrics for more complex engineering design applications.

One of the possible mechanisms of the curling in fabrics is the residual stresses of yarns~\citep{Shohreh2015}.
Mechanical energy is accumulated by bending and twisting the yarn. 
The yarn is interlaced and constrained by itself, and the contact induces moment in two directions: the wale (horizontal) direction and the course (vertical) direction~\citep{Nuray2000} (Fig.~\ref{fig1}(a)).
The internal stresses on the loop make the knit curl by binding the edge of the fabric~\citep{Hamilton1974, Quaynor1998}. 
The previous studies of fabrics uncover the mechanical and geometric aspects of knitted fabrics. Examples include a two-dimensional geometric model of the knitted loop~\citep{Chamberlain1926}, an approximation of the knitted loop length~\citep{Leaf1955}, a mechanical model based on forces acting on knitted loops in inter-loop contact~\citep{Hearle1969}, a geometric model to predict tensile properties of knitted fabrics~\citep{Chou1989}, simulation models based on yarn dynamics~\citep{Kaldor2008, Cirio2016, Cherradi2022}, and a finite element method to analyze the knitted loop shape under tension~\citep{Minapoor2015}.
However, these models have been primarily predicated on 2D analysis~\citep{Poincloux2018}, and the relationship with the 3D curling behavior has not yet been clarified so far.

Knots and knitted loops exhibit several similarities in their structural and mechanical properties. Both consist of interlaced yarns or strings that form stable configurations through their interlocking patterns. Just as the geometry and type of knot influence its mechanical properties~\cite{Ashley1944,Chisnall2020}, the specific loop structures and alignments in knitted fabrics determine their elasticity, draping characteristics, and curling tendencies~\cite{Mikuvcioniene2010,Sang2015,Singal2024}. Recent research has developed the three-dimensional elastic model of building blocks of knots called orthogonal clasp, offering important frameworks for understanding the intricate interactions within interlaced systems~\citep{Grandgeorge2021}. This model facilitates a more nuanced analysis of how individual filaments within a knot or knitted loop interact under various forces, enhancing our understanding of the mechanical behavior of knitted fabrics.

In this article, we clarify the relation between the 3D curling behavior and the shape of the knitted loop, combining experiments and simulations. We investigate the 3D structure of plain knitted fabrics with different knitting numbers in the wale and course directions ($N_w$ and $N_c$, respectively) using a commercial knitting machine, and classify curling behaviors into the following three different types: the side curl wrapped in the wale direction, the top/bottom curl wrapped in the course direction, and the double curl wrapped in both directions.
We uncover that the difference in the curling behavior is related to that in the shape of a single knitted loop. 
The different curling behavior appears through the different aspect ratios of the knitted loop geometry, which would change the internal moment along the fabric.
To simplify and investigate the 3D deformation of the knitted loop shape, we perform a compression test for half of the single knitted loop by modeling the knitting yarn as an elastic rod. The constrained rod undergoes the pop-up transition, depending on the aspect ratio of the knitted loop shape. These results imply that the change in shape per a single knitted loop is correlated with the 3D curling behavior of knitted fabrics.

\section{Fabrication of knitting samples}
\label{sec:knit_fab}

Using a commercial knitting machine (Amimumemo GK-370, Dless In, Japan) and cotton yarns (Emmy Grande, Olympus Thread, Japan), we knit rectangular-shaped
plain-knit fabric of wale (course) knitting numbers $N_w$($N_c$) defined in Fig.~\ref{fig1}(b). During knitting, a $370$ g weight is hung at the base of the knitted fabric to ensure constant proper tension. 
To prevent the knitted fabric from coming undone, the beginning and ending stitches are anchored with helically interlaced yarns (a knitting method called \textit{bind-off}).
After fabrication, each sample is placed and vibrated in an acrylic container and then conditioned in a standard atmosphere to prevent the influence of external forces as possible. In this paper, we systematically vary both the course and the wale knitting numbers as $N_c = 5-30$ and $N_w = 5-90$, respectively, and investigate the knit geometry of given $(N_w, N_c)$.

\section{Geometry of knitted fabric and single-knitted loop}
\label{sec:knit_geometry}

\subsection{Curling morphology of the plain stitches}
The knitted plain stitches of given knitting numbers $(N_w, N_c)$ are classified into three distinct states as shown in Fig.~\ref{fig1}(b): the side curl along the wale direction (circles), the top/bottom curl along the course direction (triangles), and the double curl affecting both directions (diamonds).
When the knitting aspect ratio $N_w/N_c$ is not so large as $N_w/N_c\lesssim2$, the knit curls only in the horizontal direction, curling from the front to the back (the side curl). 
As $N_c$ and $N_w/N_c$ increase, the knit curls not only in the horizontal direction but also in the vertical direction, curling from the back to the front, resulting in the double curl affecting both directions. When $N_c$ and $N_w/N_c$ increase further, the horizontal curl disappears, and consequently, the knit curls only in the vertical direction (the top/bottom curl).

\begin{figure}[h!]
        \centering
        \includegraphics[width=0.48\textwidth]{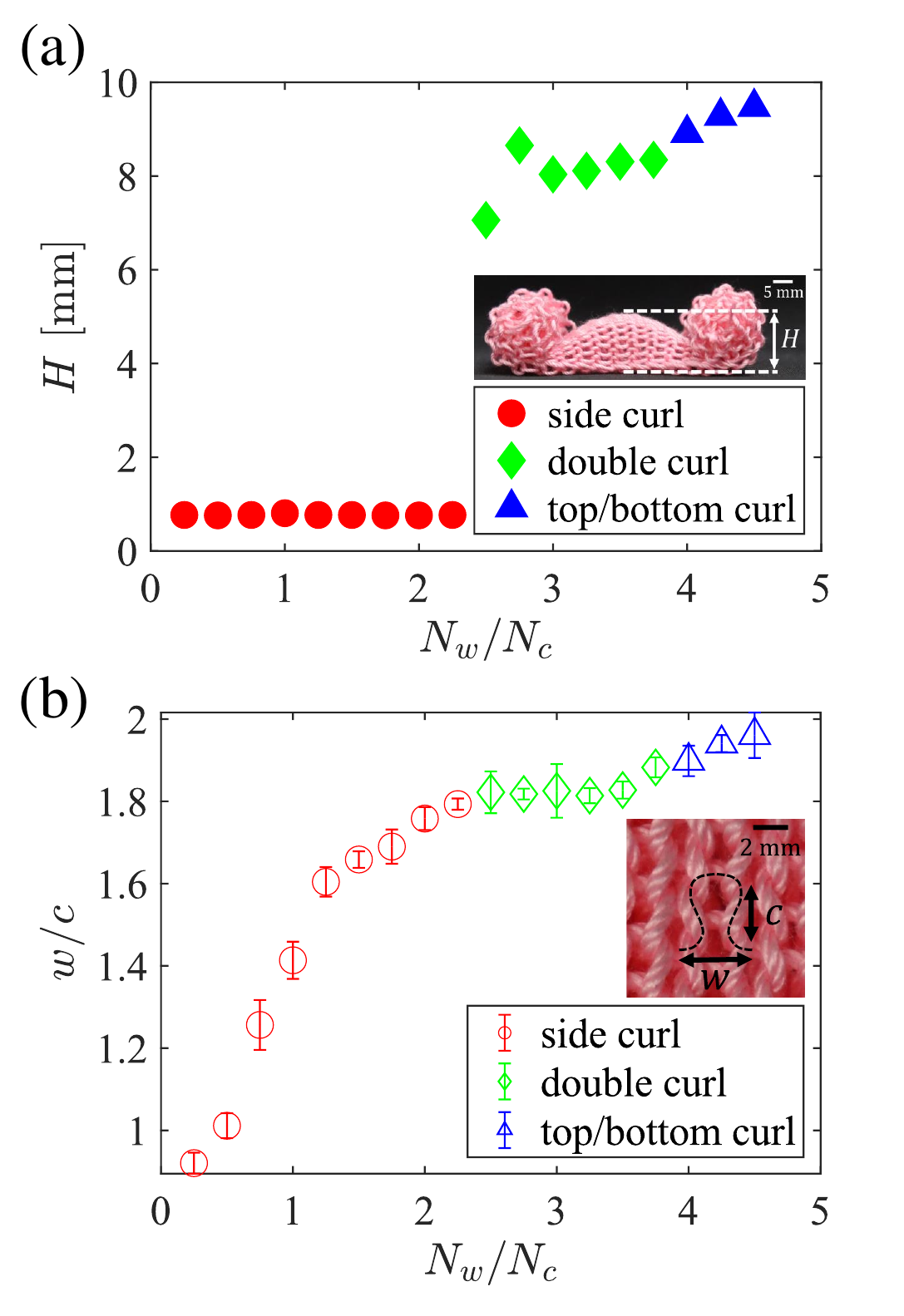}
        \caption{
        Correlation between the curling geometry and the ($\Omega$-shaped) single knitted-loop.
        (a) The curl height $H$ (defined in the inset) as a function of the aspect ratio of the number of stitches, $N_w/N_c$ for $N_c = 20$.
        (b) Relation between the aspect ratio $w/c$ of the single loop and that of the number of stitches $N_w/N_c$. The dashed line in the inset represents the shape of the typical single-knitted loop. The wale lengths $w$, and the course lengths $c$ are the characteristic lengths of the single loop defined in the inset. The colors of the data points correspond to those in Fig.~\ref{fig1}.
        }
        \label{fig2}
\end{figure}

To characterize the curling behavior quantitatively more in detail, we measure the height $H$ of the center of the fabric by image analysis (Fig.~\ref{fig2}(a)). An RGB image of the sample is converted to a binary image, and we perform edge detection with Matlab to obtain $H$.
When the aspect ratio $N_w/N_c$ is small, the height $H$ is small, indicating that the knit is side-curled. As $N_w/N_c$ becomes larger, $H$ increases, which implies that we observe double or top/bottom curls.
The combined observations of the phase diagram (Fig.~\ref{fig1}(b)) and the curling height, $H$ (Fig.~\ref{fig2}(a)) indicates that the three-dimensional shape of the plain-stitch correlates with the wale and course knitting numbers $(N_w,N_c)$.

\subsection{Geometry of the single knitted-loop}
We experimentally find that the plain stitches exhibit three different curling morphologies. Although the curling behavior is qualitatively consistent with the previous literature~\citep{Amanatides2022}, 
the detailed mechanisms have not been clear so far.
To uncover the mechanism behind the curling behavior, we focus on the shape of the single-knitted loop of each sample.

The shape of the ``$\Omega$-shaped" knitted loop is characterized by two geometric parameters, the wale length $w$ to the course length $c$, defined in the inset of Fig.~\ref{fig2}(b)~\citep{Nuray2000}. 
The wale $w$ and the course lengths $c$ are measured for six loops randomly chosen in the region with small curvature (nearly flat surface) on each sample, and obtained data through image analysis. The obtained $w$ and $c$
are averaged over the different loops.
We plot the aspect ratio of the knitted loop $w/c$ as a function of the aspect ratio of the knitting number $N_w/N_c$ in Fig.~\ref{fig2}(b). We find that as the knit shape is wider ($N_w/N_c\gg1$), the loop is horizontally elongated ($w/c$ becomes larger). 
The increase of $w/c$ is correlated with the increase of $H$ (Fig.~\ref{fig2}(a)(b)). This observation implies that $w/c$ changes according to $N_w/N_c$ and consequently affects the curling behavior from the side, double to top/bottom curls.

\section{Mechanical performance of knitted fabrics}
\label{sec:knit_mechanical_performance}

The change in the single-knitted loop shape, specifically the ratio of $w/c$, plays a significant role in the curling behavior of the knitted fabric, as discussed in Sec.~\ref{sec:knit_geometry}.
We, here, investigate the loading/unloading cycle test for the knitted fabric along both course and wale directions to comprehend how changes in loop shape affect mechanical performance. 
 A cycle test is performed using the force testing machine (EZ-LX, Shimadzu, Japan) to stretch the knitted fabric up to 30 \% of the natural length (up to $0.3$ strain), and then return to the natural length at the speed of $1$ mm/s. To reduce the effects of yarn slippage and fix the knitted fabric securely to the testing machine, we fabricate comb-shaped structures made of aluminum with $1.5$ mm uniform intervals. We hook the sides of the fabric (either along wale or course directions) with the aluminum combs and set them to the testing machine.

\begin{figure}[t!]
        \centering
        \includegraphics[width=\columnwidth]{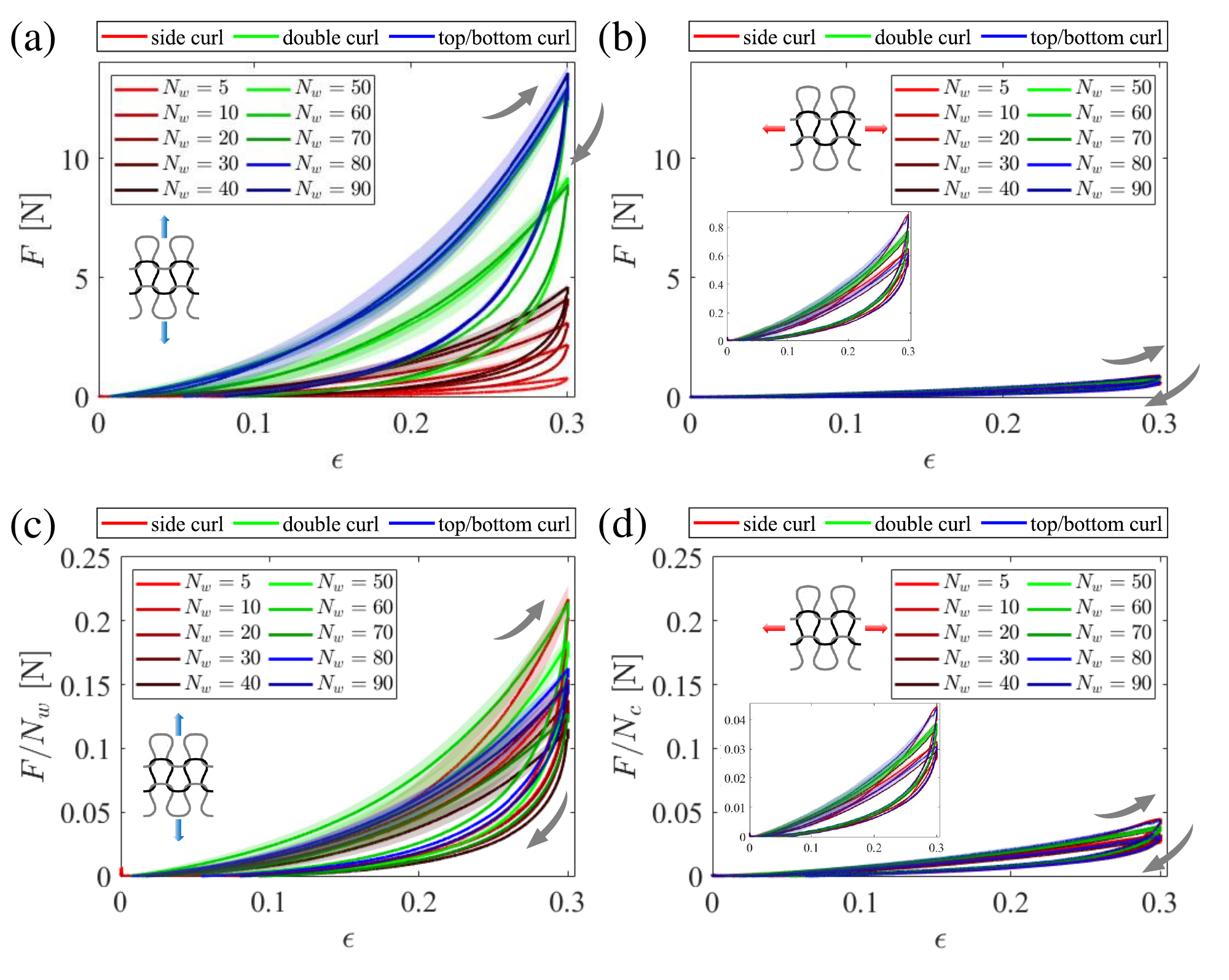}
        \caption{
        Tensile performance of plain stitches along course (a)(c) and wale directions (b)(d). 
        Force-strain curves in (a) course direction and (b) wale direction for $5\leq N_w\leq90$, while $N_c$ is fixed throughout as $N_c = 20$.
        The force-strain curves of (a) and (b) are normalized by the knitting numbers perpendicular to the corresponding tensile direction as (c)~$ F/N_w$ and (d)~$F/N_c$, respectively.
        }
        \label{fig3}
\end{figure}

The force-strain curves exhibit the characteristic hysteretic loop for both in course and wale directions, reflecting the fact that the frictional self-contact of yarn is essential for the mechanical performance of knits (Fig.~\ref{fig3}). The hysteretic force-strain curve is consistent with the previous literatures~\cite{Poincloux2018,Singal2024,Douin2023}. 
When tensile force is applied in the course direction (Fig.~\ref{fig3}(a)), the maximum forces (at $\epsilon=0.3$) increase as the curl shape changes from the side curl, the double curl, to the top/bottom curl (larger $N_w/N_c$). 
By contrast, as shown in Fig.~\ref{fig3}(b), when tensile force is applied in the wale direction, the maximum force is insensitive with the wale knitting number $N_w$.
In addition, the maximum forces along the wale direction are much smaller than those along the course direction, indicating that the knitted fabric is more stretchable along the wale than the course direction.
We have revealed that the mechanical properties of the knitted fabrics change depending on the curl shape, possibly due to the anisotropic mechanical behavior arising from the $\Omega$-shaped knitted loops. 

We now aim to examine the relation between the single-knitted loop geometry and the mechanical performance of the fabric. We normalize the force (Fig.~\ref{fig3}(a) and (b)) by the knitting numbers perpendicular to each tensile direction to examine the ``stress" per knitted loop (Fig.~\ref{fig3}(c) and (d), respectively).
We find that the dependence of the maximum force on the detailed shape of the loops is not clear. Therefore, the quantitative relationship between the detailed knitted loop shape and the overall mechanics of the knitted fabric cannot be determined within the current setup, and a more detailed study is required. Nevertheless, the findings in Fig.~\ref{fig3}(a) and (b) clearly indicate that the curl shape is related to the mechanics of the knitted fabric. In other words, there would be some relationship between the loop shape and the mechanical properties of the knitted fabric.

We have experimentally studied the geometry and mechanical performance of knitted fabrics made of yarns. We have clarified that the unit cell of the fabric, $\Omega$-shaped single knitted-loop, plays a significant role in their mechanics. From here, to understand the essential mechanisms of the fabric, we perform model system studies in two folds. In Sec.~\ref{sec:rod}, we model the single knitted loop of yarn by an elastic rod made of silicone elastomer. The elastic rod is constrained into the $\Omega$-shaped configuration, whose buckling condition is studied by combining experiments and simulations. In Sec.~\ref{sec:icd}, we simulate the whole geometry of the fabric based on the yarn-level model method proposed in the previous literature~\cite{Kaldor2008}, where the yarn is modeled by the B-spline curve with elastic stretching and bending energy. We aim to reproduce the curling shape numerically.

\section{Buckling instability of the single-knitted loop}
\label{sec:rod}

\begin{figure}[ht!]
        \centering
        \includegraphics[width=\columnwidth]{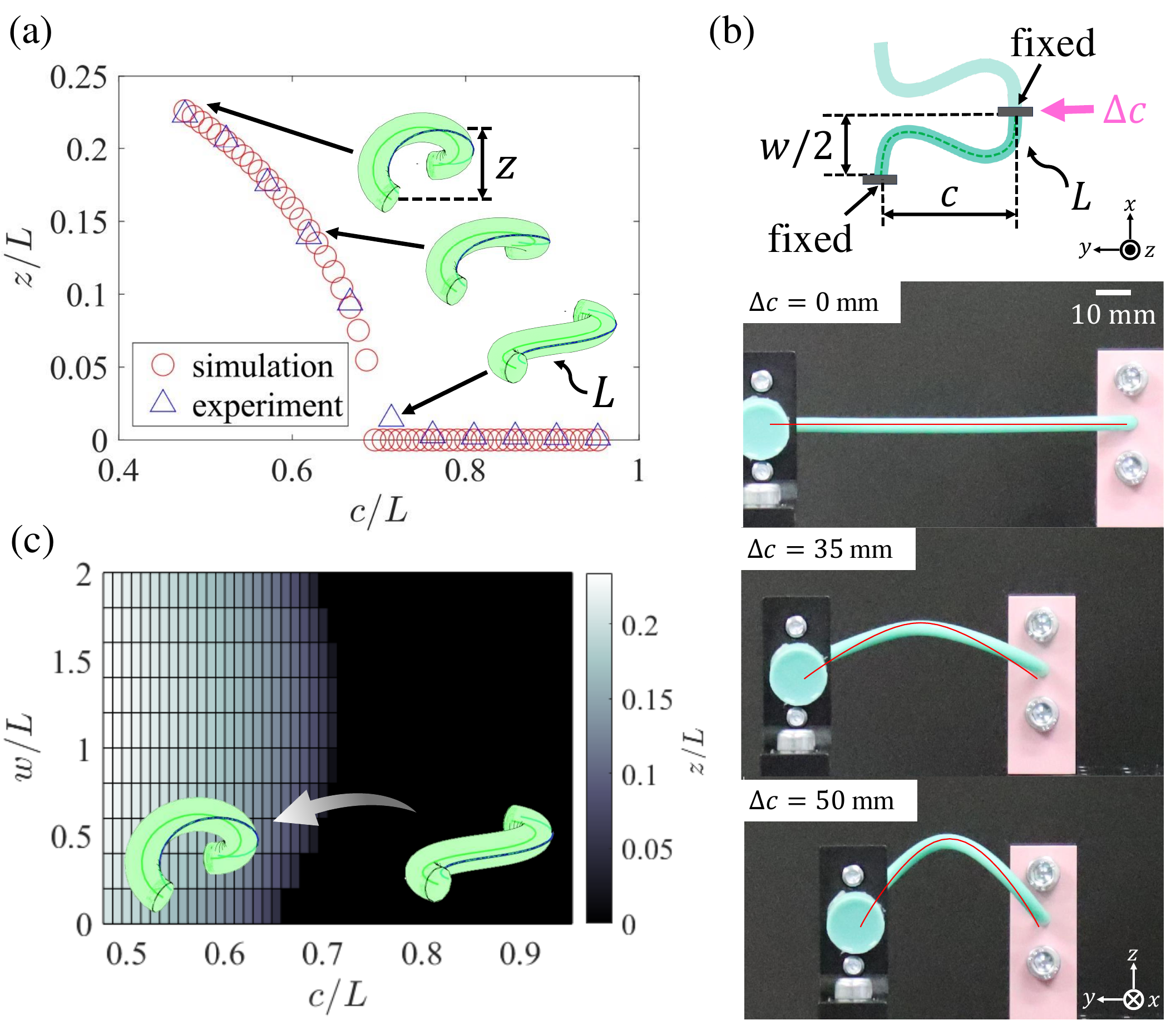}
        \caption{
        Buckling of the elastic rod model for the single-knitted loop geometry.
        (a) 
        The out-of-plane displacement of the rod-centerline, $z$, as a function of $c/L$ for $w/L=0.27$.
        (b) 
        Horizontal view of the typical buckling behavior ($x$--$z$ plane) in the experiment $(w/L = 0.27)$, where the corresponding centerline shape of the simulation is overlaid with a red solid line.
        (c) 
        The heat-map of the out-of-plane displacement, $z$, plotted on $c/L$ -- $w/L$ plane.
        }
        \label{fig4}
\end{figure}

We investigate the out-of-plane instability of loop shapes using elastic rods instead of yarns. The silicone elastomer (Elite Double 22, Zhermack, Italy) is applied to fabricate the naturally straight elastic rod. 
The rod is clamped at both ends at a length $L = 105$ mm, and is fixed to the attachment fixture in a plane S-shape so that the rod corresponds to half of the single knitted loop shape as shown in the inset of Fig.~\ref{fig4}(a). Note that the definitions of $w$ and $c$ here are identical to those for the fabric (Sec.~\ref{sec:knit_geometry}).
The height $z$ of the out-of-plane deformation of the rod is measured by fixing $w = 28.4$ mm and changing $c$ in steps of $\Delta c = 5$ mm from an initial state of 100 mm to 50 mm (Fig.~\ref{fig4}(b)). 

We analyze the mechanical equilibrium state of the elastic rod model in the same protocol as the experiment through the discrete simulation~\citep{Bergou2008,Sano2019,Korner2021,Sano2022}. 
The rod centerline is discretized into 31 connected beads, where the bend and twist energies are taken into account through the difference of Euler angles between adjacent local material frames attached to each bead~\citep{Sano2022}. The two ends of the rod are clamped as experiments. We numerically integrate the equation of motions for each bead and examine the equilibrium configuration of the centerline. Hence, there are no adjustable parameters.

We observe that the elastic rod exhibits out-of-plane buckling as the course length $c$ is varied while the wale length $w$ is fixed. The experimental data in Fig.~\ref{fig4}(a) shows that the deformation height $z$ increases as $c$ decreases at the critical course length ($c/L\simeq0.7$)\, in excellent agreement with the discrete simulation.
The buckled shape of the rod is also in excellent agreement with the experimental configurations, as shown in Fig.~\ref{fig4}(b).
In summary, both the experimental and simulation results demonstrate that the rod buckles out of the plane when the aspect ratio $w/c$ reaches a critical value. This precise concordance between experiment and simulation shown in Fig.~\ref{fig4}(a) and (b) allows us to predict the deformation of the elastic rod for a wide range of knitted loop aspect ratios $w/c$, varying both $w$ and $c$, based on our confidence in the discrete simulation (Fig.~\ref{fig4}(c)).
We find that the loop undergoes the out-of-plane deformation, depending on $w/c$. These results imply that the change in shape per a single-knitted loop is expected to be related to the curling behavior of the knitted fabrics.

\section{Reduced numerical model of the plain knitted fabric}
\label{sec:icd}

{We have shown that the aspect ratio of the single-knitted loop plays a vital role in the buckling behavior of the loop. In other words, the internal moment acting on the loop in the fabric would curl the fabric. Here, to further validate the scenario that the loop geometry triggers the curling of the knit, we perform the reduced simulation of the knitted fabric known as {yarn-level model}~\citep{Kaldor2008}, where the yarn is modeled by a naturally straight rod with the linear constitutive law. We show that our numerical modeling reproduces the experimental curling morphology qualitatively and that the loop geometry is central to determining the curling behavior of the fabric.}

\begin{figure}[h!]
        \centering
        \includegraphics[width=\columnwidth]{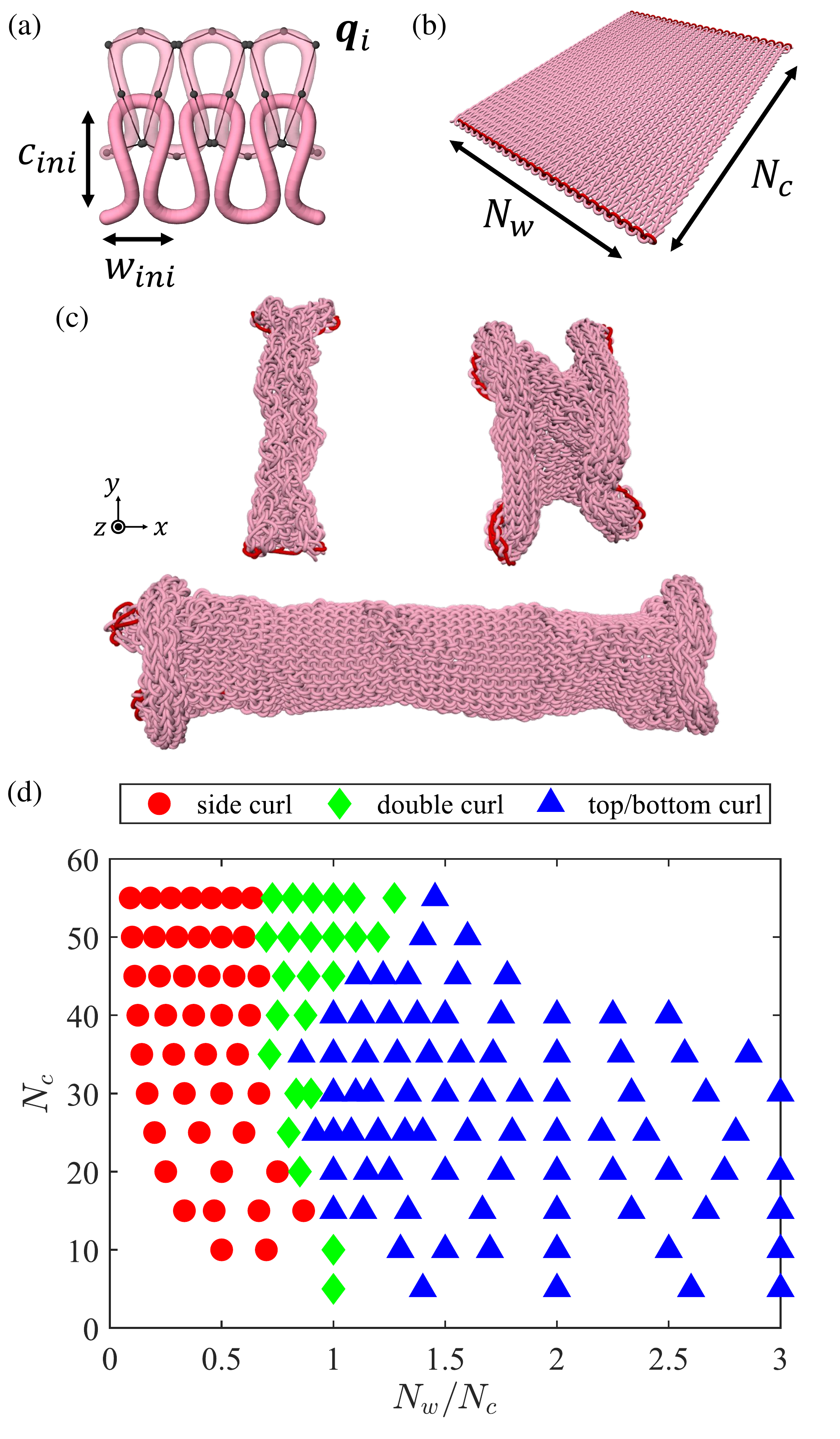}
        \caption{Reduced numerical modeling methods and results of simulations.
        (a) Schematic of the discretized yarn centerline. The continuous shape of the yarn centerline is represented by a set of smooth B-spline segments, where the position is determined by the control points $\bm{q}_i(t)$. The initial unit cell geometry, characterized by the wale ($w_{\rm ini}$
         ) and course lengths ($c_{\rm ini}$
        ), is tiled periodically to generate the knitted structure.
        (b) Initial geometry of the knitted fabric in the simulation with knitting numbers $N_w$ and $N_c$. The fabric is constructed by periodically tiling unit cells together with \textit{bind-off} (highlighted by red threads).
        (c) Typical snapshots of the three distinct curling behaviors observed in the simulation: side curl ($(N_w, N_c) = (10, 35)$), double curl ($(N_w, N_c) = (25, 35)$), and top/bottom curl ($(N_w, N_c) = (55, 35)$).
        (d) Phase diagram summarizing the classification of the final curling shapes in the simulations.
        }
        \label{fig5}
\end{figure}

{We model the shape of the bent yarn by the centerline position of the naturally straight rod, $\bm{r}(s,t)$, where $s$ represents the arclength of the centerline. The instantaneous centerline position at time $t$ is discretized into the set of $N$ number of smooth curved segments using the position of the control points, $\bm{q}_i(t)$, as
\begin{eqnarray}
    \bm{r}(s,t) = \sum_{i=1} ^{N+3} \bm{q}_i(t)b_i(s),
\end{eqnarray}
where $b_i(s)~(i=1,2,\cdots,N+3)$ represents the $i$-th B-spline curve. With the aid of the discretization, the infinite degree of freedom for the shape of the knit is reduced to the finite degree of freedom for the location of the set of control points, $\bm{q}_i(t)$~(Fig.~~\ref{fig5} (a)). 
{To derive the generalized equation of motion for $\bm{q}_i(t)$, we follow the yarn-level 1D model proposed in Ref~\cite{Kaldor2008}, except enforcing the inextensibility of segments of yarns. Hence, the stretching, bending, and contact energy functionals in terms of $\bm{q}_i(t)$ are considered to express the dynamics of yarns, where the appropriate damping terms, such as linear viscous force and normal/tangential viscous force upon contact, are added so that the elastic yarn reaches the mechanical equilibrium within the reasonable computational time. The dynamic relaxation method is employed for the time integration of the equation of motion~\cite{Wriggers2008}.}
}

{
The initial geometry of the fabric, which corresponds to the shape hanged by the weight during fabrication (Sec.~\ref{sec:knit_fab}), is generated by extending the protocol proposed in Ref.~\cite{Leaf1955}. The shape of the initial $\Omega$-shaped loop is expressed by combining the planar \textit{elastica} shape (using the elliptic integrals)~\cite{Kaldor2008} and the out-of-plane sinusoidal function of half wavelength. The unit cell geometry characterized by the initial wale, $w_{\rm ini}$, and course lengths, $c_{\rm ini}$, is periodically tiled and connected smoothly with each other to generate the knitted structure (Fig.~\ref{fig5} (a)). We then bind the loops located at the top and bottom sides of the fabric by helically-bend rods {(called \textit{bind-off})} whose self-contacts are neglected (to reduce the computational costs). See Fig.~\ref{fig5} (b) for the typical initial geometry of the fabric in our simulation. 
}

{The mechanical equilibrium shape of the fabric for a given set of $(N_w, N_c)$ in our simulation is obtained as follows. We numerically mimic the pre-stressed experimental configurations on the knitting machine {based on Ref.~\cite{Leaf1955}.} 
Then, we decrease the natural length in the stretching energy functional by the length ratio $\alpha$ to realize the pre-stretched yarn in the knit. {The pre-stretch ratio, $\alpha$, is chosen to be $\alpha = 0.2$ throughout. We confirm that the value of $\alpha$ does not affect the qualitative results below as long as $\alpha$ is small.} We remove the weight and integrate the equation of motion for the control points $\bm{q}_i(t)$ for a sufficiently long time to compute the mechanical equilibrium configuration (See Fig.~\ref{fig5}(a)).}
{We implemented the parameters in the simulation such as the diameter, bending stiffness, and other dynamical parameters identical to the previous literature~\cite{Kaldor2008}, while the Young modulus $k_\text{len}=4.00\times 10^3~\rm{ g~cm^2/s^2}$, contact force coefficient $k_\text{contact}=3.25~\rm{g/s^2}$, and the mass-proportional damping coefficient $k_\text{global}=1.00\times 10^{-2}~\rm{g/s}$ are chosen to ensure both the efficient relaxation of the pre-stretched knit and numerical stability.}

{\section{Results of the reduced numerical simulations}}

{We show typical snapshots of the three distinct curling behaviors: side, double, and top/bottom curls, reproduced through the simulation in Fig.~\ref{fig5} (c). The simulated shape is consistent with our experimental observations, where the (horizontally) wider knits curl vertically. 
We call that the knits are side-curled, if they curl horizontally only. We distinguish double and top/bottom curls based on whether the top and bottom sides of the knit are in contact or not. When they are in contact, the knit is a top/bottom curl. If not, the knit is classified as the double-curl. 
We summarize the classification of the final shape in the simulations as the phase diagram in Fig.~\ref{fig5} (d). Although the phase boundary does not match with experimental data quantitatively, the existence of three different curling shapes is consistent with the experimental results.}

\begin{figure}[h!]
        \centering
        \includegraphics[width=\columnwidth]{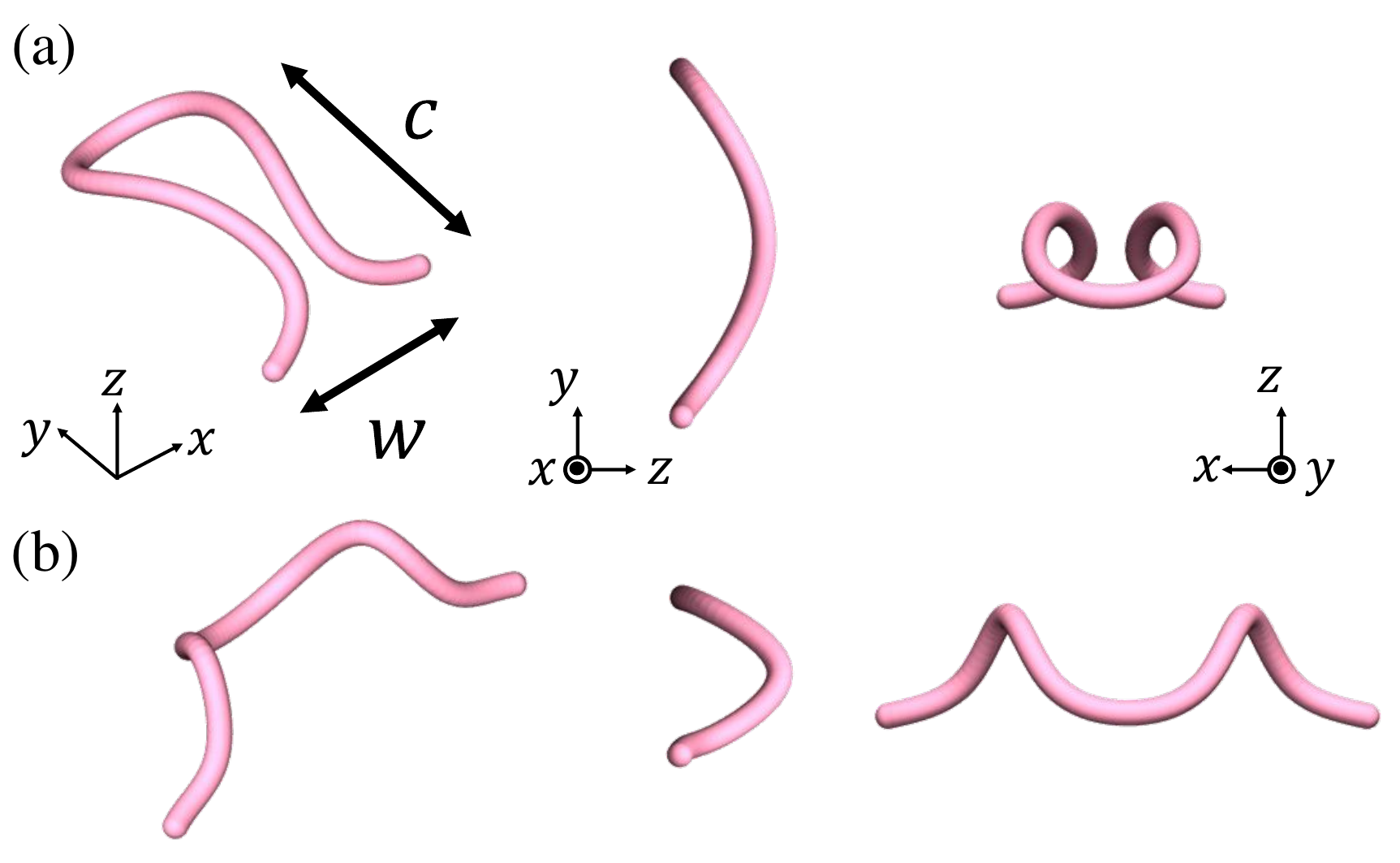}
        \caption{Comparison of the initial loop geometries.
        (a) Vertically stretched loop ($w\ll c$) where the curvature in the $z$-$x$ plane is greater than that in the $y$-$z$ plane. 
        (b) Horizontally stretched loop ($w\gg c$), exhibiting greater curvature in the 
        $y$-$z$ plane compared to the $z$-$x$ plane. The wale and course lengths ($w$ and $c$) set the characteristic length scales of the loop, influencing the curling morphology and mechanics of the fabric.
        }
        \label{fig6}
\end{figure}

{
We have shown that the simulation results based on the reduced modeling are qualitatively consistent with experiments, whereas they do not agree quantitatively. We discuss the possible two reasons below. First, we model the mechanical properties of the yarn by an elastic rod with Hookean linear spring and linear bending constitutive law. The yarn is made of several threads twisted in frictional contact with each other, realizing complex mechanical responses even for a single yarn. For example, the yarn exhibits a hysteretic response against cyclic loading/unloading protocol and will undergo plastic deformation, forming spontaneous curvatures. One may need to model the nonlinear mechanics of the yarn including its bending/twisting constitutive law and contact mechanics. Nevertheless, it is notable that three different curling behaviors are realized in simulations. Second, we model the \textit{bind-off} by helically bent rods at the top and bottom sides of the knit. Experimentally, we confirm that the stiffness of the bind-off affects the curling morphology, meaning that the boundary condition of the knit is crucial in predicting three-dimensional shapes quantitatively. 
Despite the fact that our modeling effort needs to be improved in future studies, we can successfully reproduce the curling morphology. In other words, the existence of three curling behaviors is generic and may be independent of the modeling of yarns. Indeed, we perform the same experiments using nylon threads and obtain qualitatively similar phase diagrams (The phase boundaries shift slightly compared with those of yarns). 
}

{We take advantage of our computational framework to understand the curling morphology in more detail. We prepare the initial loop shape using the elastica-shape with the sinusoidal out-of-plane displacement. The $\Omega$-shaped unit cell has three different curvatures, planar curvature on $x$-$y$ plane and two out-of-plane curvatures on $y$-$z$ (course) or $z$-$x$ (wale) planes (See Fig.~\ref{fig6}). We compare the initial geometry of the horizontally ($w\ll c$) and vertically ($w\gg c$) stretched loops in Fig.~\ref{fig6} (a)(b). The wale and course lengths, $w$ and $c$, set the characteristic length scales of the loop and the stored bending energy of the loop. When the knit relaxes to the force-free state, the bending energies of the initial loop, such as those along the wale and course estimated as $\sim1/w$ or $\sim1/c$, respectively, are distributed along the knit. The knit would curl along the direction whose bending energy is reduced more efficiently. The ratio of characteristic bending energy is estimated as $w/c$, indicating that the aspect ratio of the loop is crucial in the curling morphology. In other words, the three-dimensionality of the single loop matters the 3D curling shape and mechanics of the fabric. }

\vspace{-0.3cm}
\section{Conclusion}
\label{sec:conclusions}
\vspace{-0.3cm}
In this study, we have investigated the geometry and mechanical performance of knitted fabrics, focusing on the 3D curling behavior and its correlation with the knitted loop shape. Our experimental analysis has revealed that the curling behavior of these fabrics is categorized into three distinct states: side curl in the wale direction, top/bottom curl in the course direction, and double curl affecting both directions. We demonstrate that these curling behaviors are influenced by the wale and course knitting numbers ($N_w$, $N_c$).
We find that the curling behavior is related to the local structure of the fabric --geometry of a single-knitted loop-- by measuring the curling height $H$.
Mechanical testing through loading/unloading cycles reveals the anisotropic mechanical properties of the knitted fabrics, indicating a relationship between the curling behavior and the overall fabric mechanics.  
To understand the fundamental mechanisms behind these observations, we model the single-knitted loop using an elastic rod made of silicone elastomer and simulate its buckling behavior. Both experimental and simulation results show that the rod undergoes out-of-plane buckling when the loop aspect ratio $w/c$ reaches a critical value, confirming the curling behavior observed in the knitted fabrics.
{In the reduced simulation of the knit, we show that our computational model can reproduce the experimentally observed curling shapes qualitatively. Although more modeling efforts are required for their further quantitative agreement, we find that the existence of three curling behaviors, side, double, and top/bottom curls, are generic in the plain stitch.}

In conclusion, our study highlights the crucial role of the single-knitted loop in determining the geometry and mechanical performance of knitted fabrics. The insights gained from this work lay the foundation for future studies aimed at uncovering the detailed mechanisms of 3D fabric deformation and improving the design and performance of knitted materials. Further research, particularly in the quantitative relationship between loop shape and fabric mechanics, is essential to fully understand and apply these findings in practical applications, such as wearable devices, soft actuators, and the design of composite materials.

\noindent \textbf{Acknowledgments.} This work was supported by MEXT KAKENHI 24H00299 (T.G.S.), 23K17317 (R.T.), JST FOREST Program, Grant Number JPMJFR212W (T.G.S.). We thank Hokuto Nagatakiya for his discussion in the initial stage of this research. 

\vspace{-0.3cm}
\bibliographystyle{elsarticle-num-names}
\bibliography{sn-bibliography}

\end{document}